\documentclass[a4paper,11pt]{jpconf}
\usepackage{graphicx}
\usepackage{amsmath}
\begin{document}
\title{Irradiation study of UV Silicon Photomultipliers for the Mu2e Calorimeter}
\author{ S Baccaro$^{1}$,  A Cemmi$^{1}$, M Cordelli$^{2}$, E Diociaiuti*$^{2,3}$, R Donghia*$^{2,3}$, A~Ferrai$^{4}$, S~Giovannella$^{2}$, S Miscetti$^{2}$, S~M\"uller$^{4}$, M Pillon$^{5}$ and I Sarra$^{2}$}

\address{ $^1$ ENEA UTTMAT-IRR, Casaccia R.C., Rome, Italy}
\address{$^2$ Laboratori Nazionali di Frascati - INFN, Frascati, Italy}
\address{$^3$ Universit\'a degli Studi Roma Tre, Rome, Italy}
\address{$^4$ HZDR, Helmholtz-Zentrum Dresden-Rossendorf, Germany}
\address{$^5$ ENEA FNG, Frascati, Italy}

\ead{eleonora.diociaiuti@lnf.infn.it,raffaella.donghia@lnf.infn.it}

\begin{abstract}
The Mu2e calorimeter is composed of 1400 un-doped CsI crystals, coupled to large area UV extended Silicon Photomultipliers (SiPMs), arranged in two annular disks. This calorimeter has to provide precise information on energy, timing and position resolutions. It should also be fast enough to handle the high rate background and it must operate and survive in the high radiation environment. Simulation studies estimated that, in the highest irradiated regions, each photo-sensor will absorb a dose of 20 krad and will be exposed to a neutron fluency of $5.5\times10^{11}$n$_{1MeV}$/cm$^2$ in three years of running, with a safety factor of 3 included.
At the end of 2015, we have concluded an irradiation campaign at the Frascati Neutron Generator (FNG, Frascati, Italy) measuring the response of two different 16 array models from Hamamatsu, which differ for the protection windows and a SiPM from FBK. In 2016, we have carried out two additional irradiation campaigns with neutrons and photons at the  Helmholtz-Zentrum Dresden-Rossendorf (HZDR, Dresden, Germany) and at the Calliope gamma irradiation facility at ENEA-Casaccia, respectively.\\
A negligible increment of the leakage current and no gain change have been observed with the dose irradiation. On the other hand, at the end of the neutron irradiation, the gain does not show large changes whilst the leakage current increases by around a factor of 2000. In these conditions, the too high leakage current makes problematic to bias the SiPMs, thus requiring to cool them down to a running temperature of $\sim0~^{\circ}$C.
\end{abstract}

\section{The Mu2e experiment}
\label{sec:Mu2e}
\begin{figure}[h!]
\centering
\includegraphics[width=\textwidth]{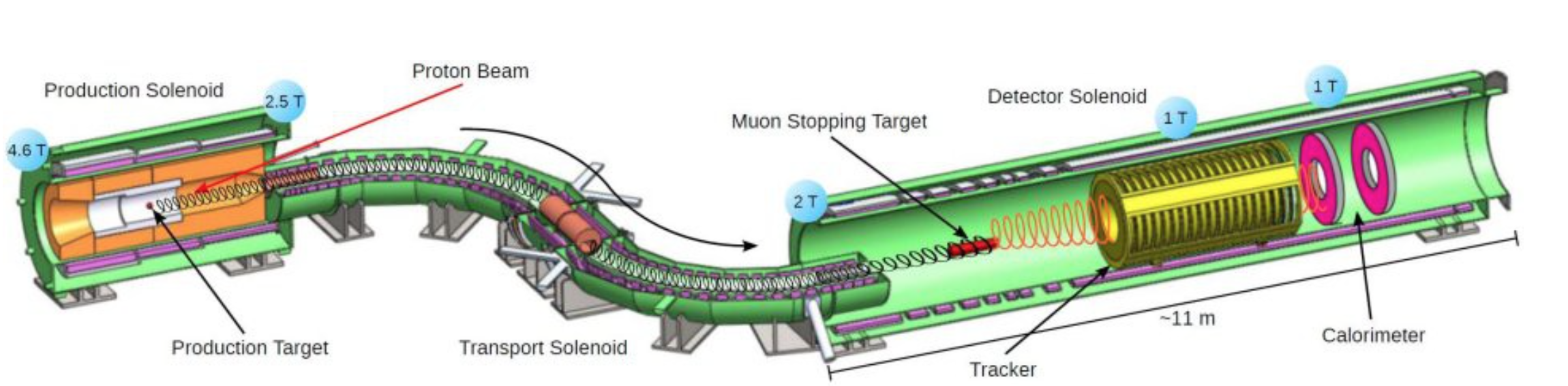}
\caption{Schematic view of the experimental apparatus.\label{fig:mu2e_marmittone}}
\end{figure}

The goal of the Mu2e experiment  is to search for  the neutrinoless, coherent conversion of muons into electrons in the field of a nucleus and improve by four orders of magnitude the previous sensitivity set at 90 $\%$ C.L. by the SINDRUM II experiment \cite{sindrum}.
This corresponds to a limit on the ratio between the conversion and nuclear muon capture rates $R_{\mu e}$ of : $(R_{\mu e}< 6 \times 10^{-17}) $, Mu2e hopes to achieve a Single Event Sensitivity  (SES) of $ 2.5\times 10^{-17}$ with $\sim 0.5$  event background.
The Mu2e apparatus is extensively documented in its Technical Design Report \cite{TDR}. As shown in \figurename~\ref{fig:mu2e_marmittone}, the layout of the Superconducting Solenoid Magnet System has a typical S-shape. In order to limit backgrounds from muons that might stop on gas atoms and to reduce the contribution of multiple scattering for low momentum particles, the inner bore of the solenoids is evacuated to $10^{-4}$ Torr.
The solenoids are organized into three sub-systems: Production Solenoid (PS), Transport Solenoid (TS) and Detector Solenoid (DS).
The 8 GeV proton beam (i.e. above the antiproton production threshold energy) coming from the Fermilab accelerator system enters the PS, hitting the production target. The reaction products, of selected charge sign, are transported through the S-shaped Transport Solenoid, which is long enough to allow the decay of almost all hadrons while suppressing line-of-sight particles. The resulting negative muon beam enters the Detector Solenoid and hits the aluminum target: $\sim50\%$ of the muons are stopped at a frequency of $\sim10$ GHz while the other $\sim50\%$ proceed toward the muon beam stop. $40 \%$ of the muons stopped are captured by the atoms and decay or convert into electrons while the rest are captured by the nucleus ($60\%$). Momentum and energy of the electrons produced in the Decay in Orbit (DIO) and Conversion Electrons (CE) events are measured by  the cylindrical-shaped tracker and by the two-disks calorimeter, respectively.
Downstream of the proton beam pipe, outside the PS, an extinction monitor is used to measure the number of protons in between two subsequent proton pulses. The Detector Solenoid is surrounded by a cosmic ray veto system. Outside the DS, a stopping target monitor is used to measure the total number of muon captures. 
\subsection{The Mu2e calorimeter}
\begin{figure}[h!]
 \centering
 \begin{minipage}[c]{.3\textwidth}
 \centering
 \includegraphics[width=\textwidth]{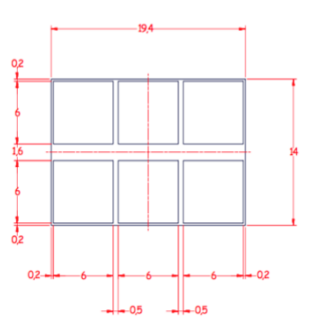}
 \end{minipage}
 \hspace{9mm}
 \begin{minipage}[c]{.4\textwidth}
 \centering
 \includegraphics[width=\textwidth]{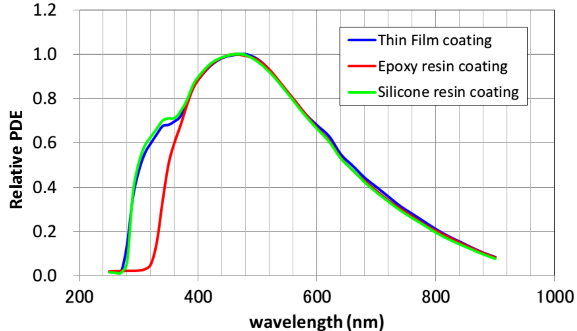}
 \end{minipage}
 \caption{Left: Sketch and schematic structure (from left to right) of the front, cross and back side of the  Mu2e SiPMs. Right: PDE for different models of Hamamatsu SiPMs. \label{fig:sipm}}
 \end{figure}

\begin{figure}[h!!]
 \centering
 \includegraphics[width=.8\textwidth]{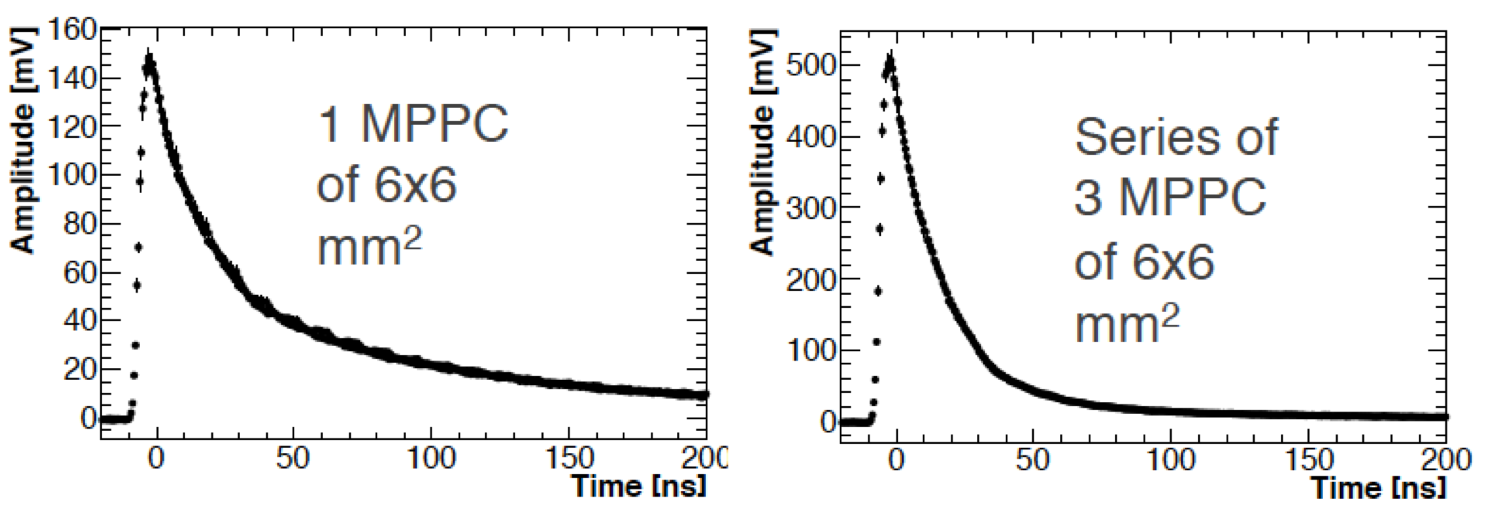}
 \caption{Left: Signal width of a single $6\times 6$ mm$^2$ SiPM. Right: Signal width of a series of three $6\times 6$ mm$^2$ SiPMs.\label{fig:confrontosipm}}
\end{figure}

The request of the Mu2e calorimeter are to provide: (i)  shower shape, energy, and timing information that, in combination with information from the tracker, can distinguish electrons from muons and pions,  (ii)  a ``seed'' to improve tracker pattern recognition and reconstruction efficiency, (iii) the means to implement an independent trigger based on the sum and pattern of
energy deposition. Moreover, the Mu2e calorimeter must operate in a high-rate, high-radiation environment, in vacuum and in presence of a 1T magnetic field. This motivates a fast response, an excellent time resolution and good radiation hardness. After a long R$\&$D phase, the best compromise between costs and properties has been selected: the calorimeter design consists of 1346 un-doped CsI crystals located down-stream of the tracker, arranged in two disks, positioned at a distance of half wavelength  of a typical conversion electron. For an overall description of the calorimeter, see \cite{DiFalco}. \\
The crystals are parallelepipeds with squared faces of $(34 \times 34)$ mm$^2$ dimensions  and are 200 mm long. Each crystal is read by two $2 \times3 $  arrays of individual $ 6\times 6$ mm $^2$  UV-extended Silicon Photomultipliers (SiPMs), see \figurename~\ref{fig:sipm} left. The solid-state photodetectors are necessary due to the presence of the high magnetic field. 
In order to match the wavelength of the emitted light produced by the CsI crystals (that is centered at 315 nm) the SiPMs have to be extended in the UV region, as shown in \figurename~\ref{fig:sipm} right.
The serial connection has been chosen  to overcome the issues related to the parallel connection which might affect the energy and time measurements, owing to the very large capacitance resulting in an increased noise, signal rise time and width. This is shown in \figurename~\ref{fig:confrontosipm}, where the signal width of a single SiPM and the series of three SiPMs are displayed.

Differently from the parallel configuration, where the signal becomes wider, the pulse shape of a series of SiPMs results narrower than that of a single SiPM. This is due to the reduction of the total capacitance of the circuit. The shorter signal decay time  minimizes the overall width of the Crystal+SiPM pulse height thus improving the pileup discrimination capability.\\
The Front End chips are connected directly to the sensors while the  Slow Control and Digitization boards are mounted on crates located on top of each disk. The electronics must work adequately in vacuum and in presence of a high magnetic field and high radiation environment.\\
The equalization of the crystal response will be provided through a circulating radioactive source (Fluorinert$^{TM}$, C$_8$F$_{18}$), already experimented by BaBar \cite{Sorgente} while a laser flasher system will be used for relative calibration and gain monitoring of the SiPMs. Usage of cosmic ray  and DIO events for calibrating "in-situ" along running is also planned.
%%%%%%%%%%%%%%%%%%%%%%
\section{ The SiPM irradiation campaigns}
\begin{figure}[h!!]
 \centering
 \includegraphics[width=.6\textwidth]{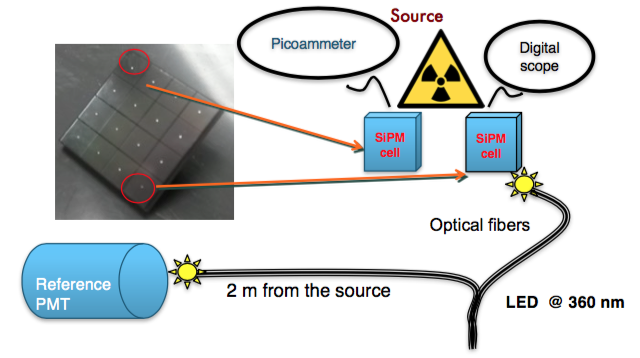}
 \caption{General scheme of the experimental setup used during the 2015/2016 irradiation campaigns.\label{fig:irrag_setup}}
\end{figure}

%%%%%%%%%%%%%%%%%%%%%%
The Mu2e calorimeter must operate and survive in a high radiation environment. 
Simulation studies estimated that the Mu2e SiPMs have to withstand an equivalent neutron fluency, at 1 MeV energy, of $4\times10^{11}$~n/cm$^{2}$ and  a  Total Ionizing Dose (TID) of 20 krad of photons ~\cite{MCirr}. These values assumes three years of running and a factor of 2 safety as calculated in the hottest irradiated region, i.e. in the innermost ring of the first disk. For this reason, we have tested our sensors before and after irradiation to measure the variation of the leakage current and of their gain.\\
During the first  irradiation campaigns in 2015, different models of UV-extended SiPMs have been tested: two from Hamamatsu \cite{Hamamatsu}, made of 16 $3\times3$ mm$^2$ cells, and one from FBK \cite{FBK}. The two Hamamatsu SiPMs had equal layout but  different protection material: one  had a protection window with a silicon protection layer (SPL) while the other one had a Micro-Film (MF). The FBK SiPM was instead a 6$\times$6 mm$^2$ monolithic.
 The general scheme of the experimental set-up used is shown in \figurename~\ref{fig:irrag_setup}.  
For the Hamamatsu SiPMs,  we acquired the signal of two cells:
 one was used to read the leakage current, while the other one provided the response to a fixed UV led pulse, i.e. the gain. In order to precisely  monitor  the input led light, a UV photomultiplier has been illuminated with the same led pulse by means of a split fiber and placed in a safe region located 2 m far away from the source.

 For the monolithic FBK SiPM, only the leakage current has been measured.
Finally during 2016 a monolithic $6 \times 6$ mm$^2$ SiPM, UV extended by SPL type, has been tested at HZDR using a similar  setup.
In the latter case, both the leakage current and the response to a UV led have been collected.
%%%%%%%%%%%%%%%%%
\subsection{Dose irradiation test}
%%%%%%%%%%%%%%%%%%
Irradiation tests with a ionization dose have been performed at the 
ENEA CALLIOPE facility~\cite{CALLIOPE}, where a $^{60}$Co source 
produces $\gamma$'s with an energy of 1.25 MeV. The activity of the source during our tests was 
$0.35\times 10^{15}$~Bq, allowing to reach from 2 to 10 Gy/h at about 
5 m distance.\\
The SPL SiPM was irradiated with these photons for three days absorbing a total dose of $\sim$ 20 krad.
The dose effect on SiPM performances is negligible both in term of leakage current and signal amplitude. 
\begin{figure}[h!]
\begin{minipage}[c]{\textwidth}
\centering
\includegraphics[width=.5\textwidth]{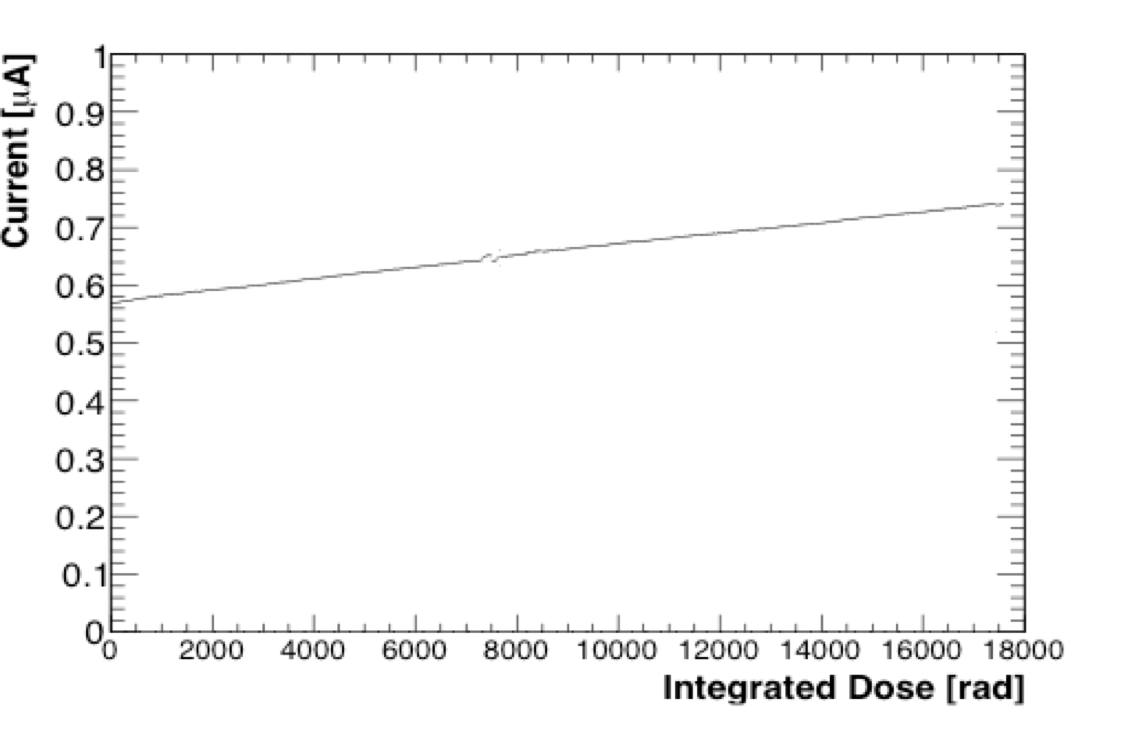}
%\caption{Leakage current of a SPL SiPM as a function of the integrated dose.\label{fig:boh}}
\end{minipage}
\hspace{9mm}
\begin{minipage}[c]{\textwidth}
\centering
\includegraphics[width=.7\textwidth]{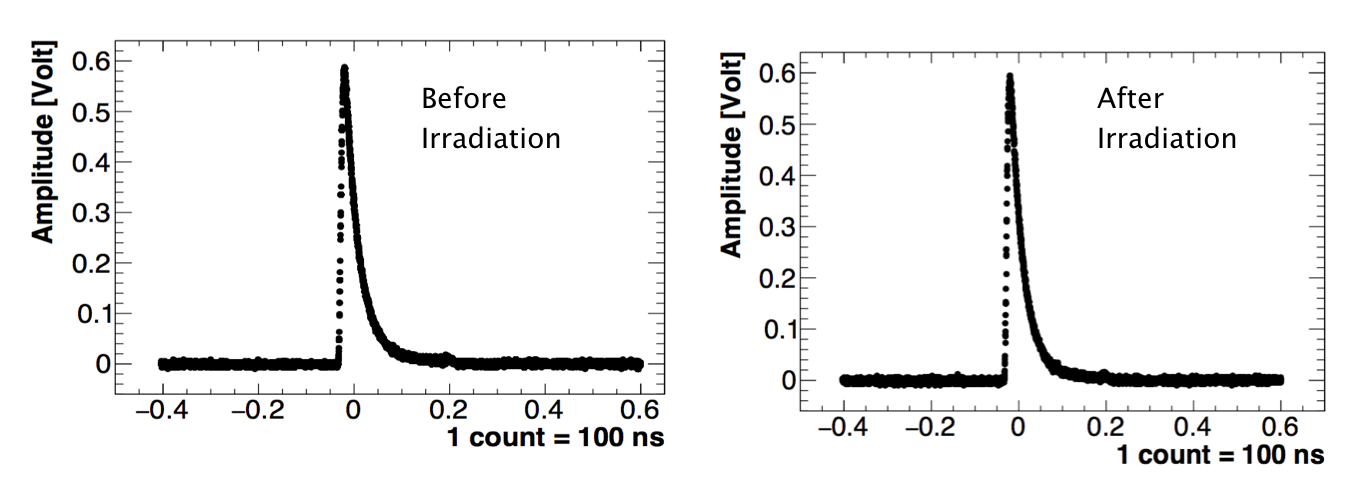}
\end{minipage}
\caption{Top: leakage current of a SPL SiPM as a function of the integrated dose. Bottom: SiPM amplitude before (left) and after (right) the dose irradiation.\label{fig:boh}}
\end{figure}
As shown in Fig.~\ref{fig:boh}  the leakage current, which before the irradiation was of $\sim 0.15~\mu$A, increased to $\sim 0.6~\mu$A as soon as the irradiation started due to the Compton effect on the SiPM active surface. In three days of irradiation the current increased by $0.15 ~\mu$A, thus practically doubling the initial dark current. The signal amplitude remained unchanged.

%%%%%%%%%%%%%%%%%%%%%
\subsection{ Neutron irradiation tests}
\begin{figure}[h!]
\centering
\includegraphics[width=.5\textwidth]{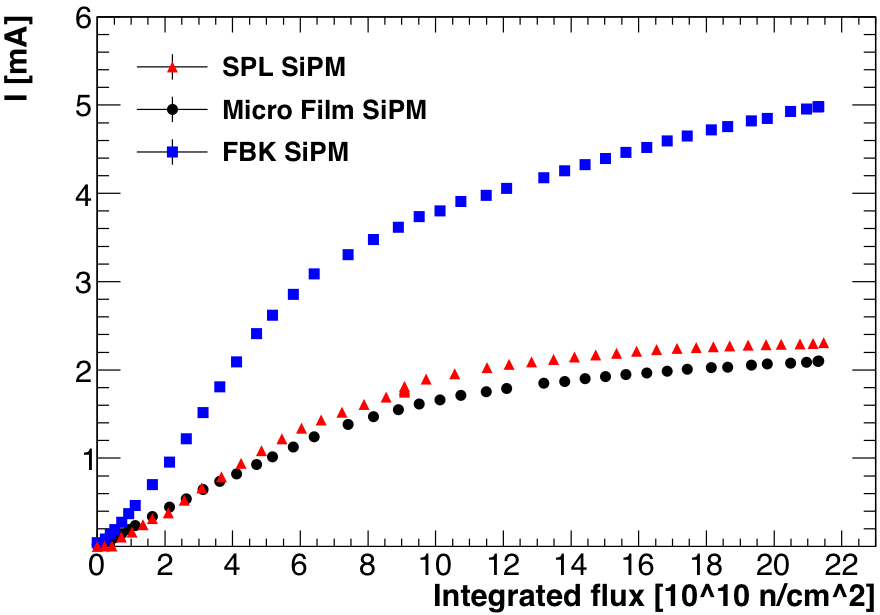}
\caption{Leakage current of a SPL SiPM as a function of the integrated dose. Results are reported for the different SiPM typologies tested. \label{Fig:I_flux_neutron}}
\end{figure}

%%%%%%%%%%%%%%%%%%%%%
The 2015 neutron irradiation tests have been performed at the ENEA 
FNG facility \cite{FNG}, in Frascati, where a nearly isotropic 14 MeV neutron flux is produced
by stopping an accelerated deuteron beam on a Tritium target. 
The maximum neutron intensity is $0.5\times 10^{11}$~n/s, close to 
the target, with a 4$\pi$ uniform production and a dependence on the distance, 
$R$, as $1/R^2$. The desired neutron intensity is reached by either 
positioning the SiPMs at the needed distance or by changing the deuteron
beam intensity. The temperature of the experimental hall was monitored and maintained between $20~^{\circ}$C and $25~^{\circ}$C.\\
The three different SiPM typologies, described before, were placed 7 cm far away from the source and  exposed, in less than 4 hours, to a fluency of $2.2 \times 10^{11}$ n/cm$^2$, which corresponds to a neutron fluency  equal to 2.2 times  that expected in the experimental lifetime.The signal peak of the SPL SiPM decreased from $\sim$250 mV to $\sim$30 mV while the MF one  decreased from $\sim400$~mV to $\sim50$~mV. The leakage current of the different SiPMs tested, as a function of the integrated flux, is reported in Fig.~\ref{Fig:I_flux_neutron}. 
In order to compare results from  SiPMs of different cell dimension, the FBK current has been reduced by a factor of 4 due to its larger active area.\\ 
A rising behavior of the dark current is clearly visible on all SiPMs: the leakage current of the MF SiPM increased from 16$\mu$A to 2mA, the one of the SPL SiPM from 100~$\mu$A to 2.2~mA and the FBK one from  $\sim21~\mu$A to $\sim 5$~mA. The observed response to the led light decreased of more than a factor of 3. Even if the hall temperature was quite stable during irradiation, the  observed gain drop was dominated by the SiPM temperature increment due to the Joule effect.

\begin{figure}[h!]
 \centering
 \includegraphics[width=.4\textwidth]{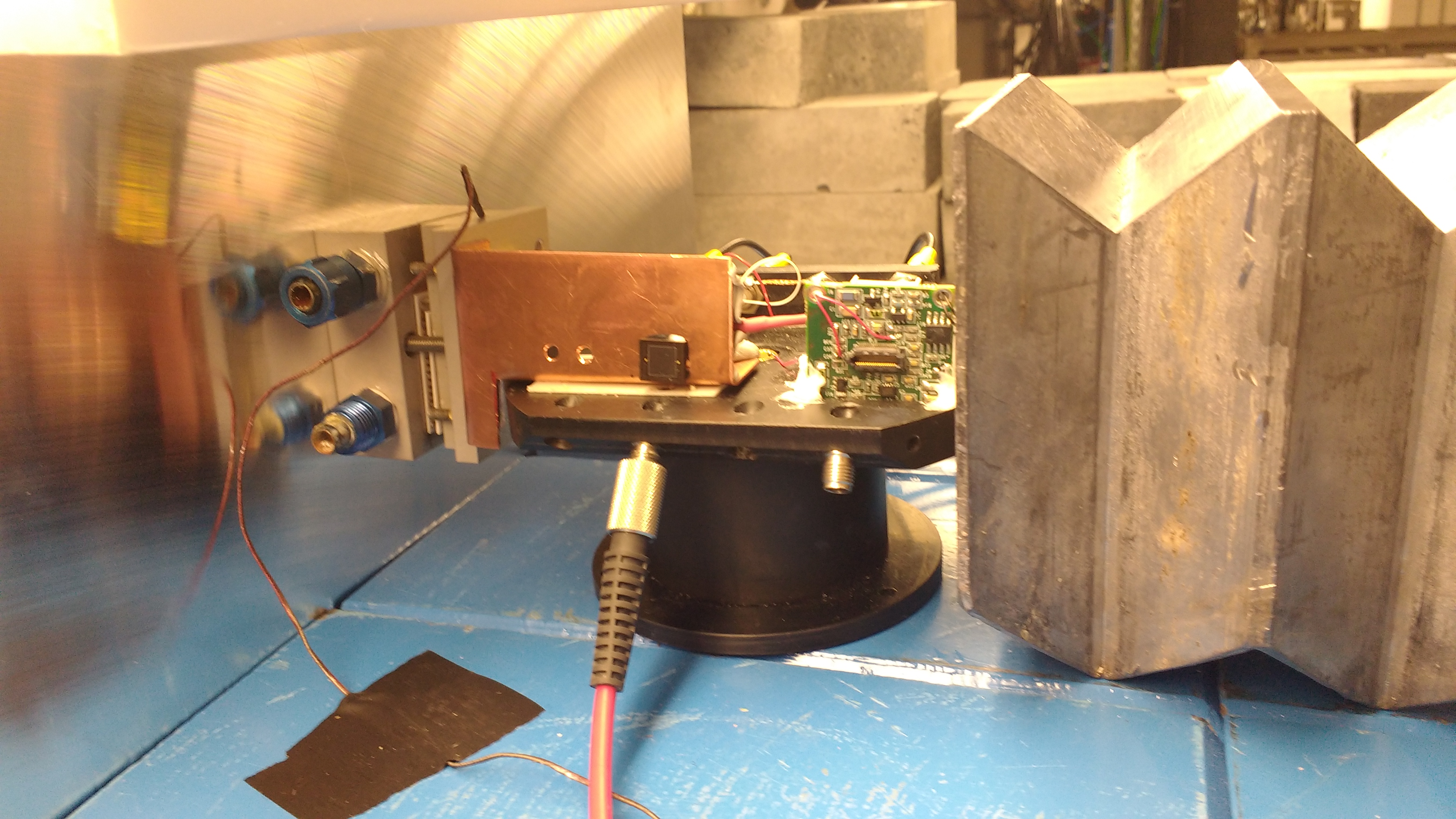}
 \caption{\emph{Experimental set up used at HZDR center.}}
\label{fig:dresda_set_up}
  \end{figure}

\begin{figure}[h!]
 \centering
 \includegraphics[width=.6\textwidth]{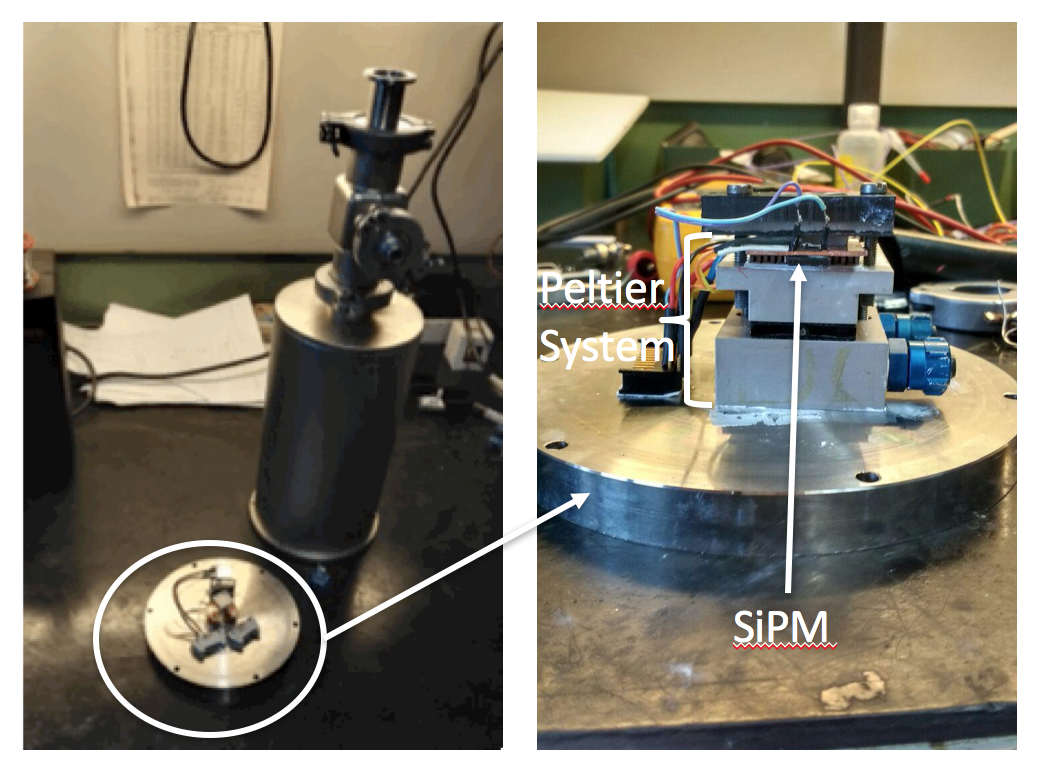}
 \caption{\emph{Experimental set up used for the temperature dependency studies. }}
 \label{fig:bombolotto_T}
  \end{figure}

\begin{figure}[h!!]
 \centering
 \includegraphics[width=\textwidth]{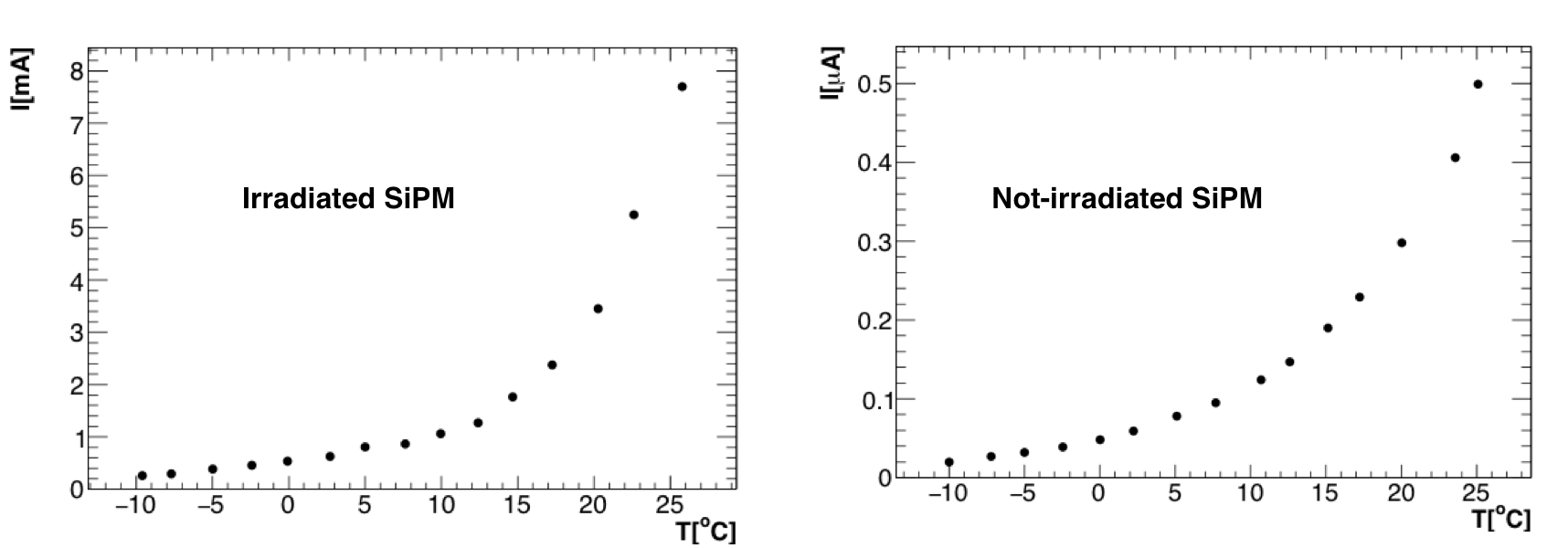}
 \caption{\emph{ Dependency of dark current on device temperature for a irradiated and un-irradiated SiPM.}}
\label{fig:I_T}
  \end{figure}

In order to confirm this, during 2016,  an independent  irradiation test has been performed at the  EPOS  source (HZDR, Dresden) with 1 MeV neutrons. The experimental setup used is reported in \figurename~\ref{fig:dresda_set_up}: a SPL SiPM  was located over the source in a place where no dose was present and with the active area positioned parallel to the incoming neutron flux. The SiPM was monitored both in current and in charge while illuminated with an UV led. To maintain the SiPM temperature as stable as possible, the irradiated SiPM was connected to a Peltier cell, with the hot side glued to a cooling system. The SiPM temperature was monitored using a PT1000. To control and monitor the Hamamatsu devices, the same experimental setup reported in \figurename~\ref{fig:irrag_setup} was used. The SiPM was biased at $V_{op}=54.9$~V.
The total neutron fluence absorbed by the SiPM in five days was $\sim 5.5~\times 10^{11}$ n$_{\rm 1 MeV}$/cm$^2$, that  is  three times the flux expected in the hottest region in 3 years of running.
 \begin{figure}[h!!]
 \centering
 \includegraphics[width=.8\textwidth]{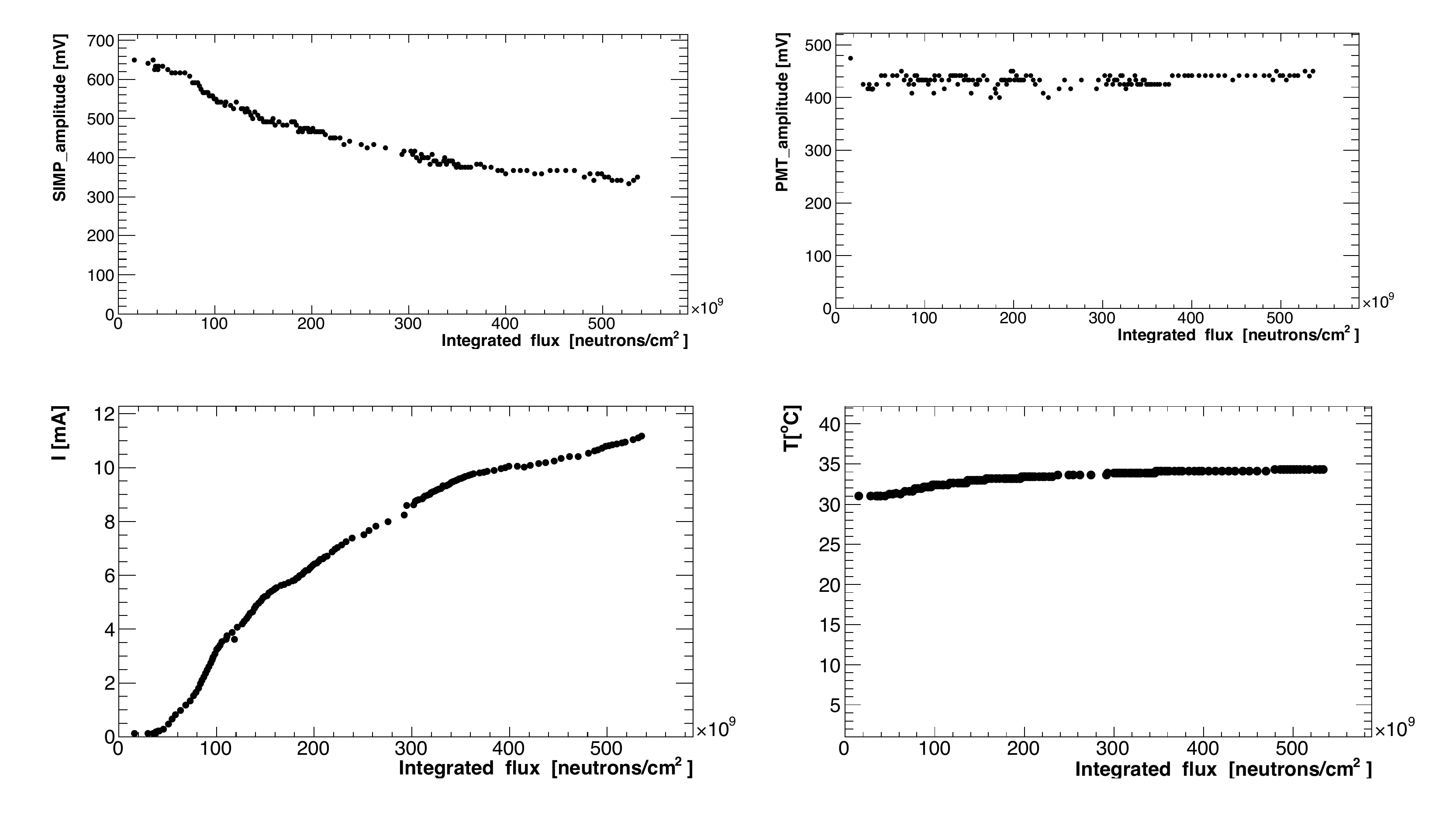}
 \caption{\emph{Top: variation of SiPM (left) and (PMT) signal amplitude during the irradiation test. Bottom: SiPM leakeage current (left) and SiPM temperature (right) as a function of the integrated flux.}}
\label{fig:risultati_dresda}
  \end{figure}
In \figurename~\ref{fig:risultati_dresda}, the  irradiation results are shown as a function of the integrated flux. 
The signal peak decreased from $\sim$650~mV to $\sim$400~mV, due to the residual
temperature variation. A rising behaviour  of the SiPM leakage current, from 60$~\mu$A  up to 12~mA, is instead clearly visible. \\
 In order to study the SiPM properties as a function of its temperature, the SiPM irradiated at EPOS and one unirradiated SiPM have been tested in a vacuum chamber at $\sim 10^{-4}$~mbar and cooled by means of a cascade of two Peltier cells. The SiPM temperature was monitored by a PT1000 sensor. 
A UV- LED was  illuminating directly the sample inside the chamber. The experimental set up used is shown in \figurename~\ref{fig:bombolotto_T}.\\
  The dark current was recorded and plotted as a function of the temperature using a pico-ammeter. In order to maintain the gain constant during the test, the SiPM was illuminated with a LED and the peak of the signal pulse height acquired with a digital scope. When varying the temperature, we adjusted the operation voltage by keeping constant the signal amplitude.   Results are in \figurename~\ref{fig:I_T}. The shape of the two distributions is similar but the dark current for the irradiated SiPM is larger by at least three orders of magnitude.
The bias supply of the front-end electronics requires the current of each channel to be $<2$~mA so that it is necessary to run the SiPM at T $<0~^{\circ}$C. 
Since, as shown in \figurename~\ref{fig:I_T}, at $0~ ^{\circ}$C the current of a single $6 \times 6$ mm$^2$ SiPM reaches $\sim1$~mA, and a serial connection  of three $6\times 6$ cm$^2$ SiPMs will have the same current, we expect that in the parallel configuration of the two series a current I $<2$~mA will flow,  satisfying our requirements.\\
At the moment of writing, the Mu2e calorimeter group is designing the front end electronics and the sensor cooling system to keep the SiPM 
at a running temperature between $-10~^{\circ}$C and $0~^{\circ}$C.
%%%%%%%%%%%%%
\section{Conclusions}
%%%%%%%%%%%%%
The determination of the SiPM charge and leakage current variation during the irradiation tests provided an important benchmark for the Mu2e calorimeter where a high radiation environment is foreseen. While the total neutron flux causes a large increase of the leakage current, a dose up to 200 Gy causes a negligible effect. The response is slightly affected by the irradiation. Changes are still acceptable for the running condition in the experiment but this requires to cool down the SiPMs to a running temperature of $\sim 0~^{\circ}$C.
\section*{Acknowledgments}
This work was supported by the EU Horizon 2020 Research and Innovation Programme under the
Marie Sklodowska-Curie Grant Agreement No. 690835.


\begin{thebibliography}{99}
\bibitem{sindrum}
W. Bertl et al. \emph{A search for $\mu-e$ conversion in muonic gold}. EPJ C, 47(2):337-346, 2006
\bibitem{TDR}
L. Bartoszek et al., \emph{Mu2e Technical Design Report}, arXiv:1501.05241
\bibitem{DiFalco} 
N. Atanov et al.,\emph{The Calorimeter of the Mu2e experiment at Fermilab}, these proceedings.
\bibitem{Sorgente}
B. Lewandowski. \emph{The BaBar electromagnetic calorimeter}. Nucl. Instrum. Meth., A494(1-3):303  -307, 2002. Proc. 8th International Conference on
Instrumentation for Colliding Beam Physics.
\bibitem{MCirr}
  G.\ Pezzullo and B.\ Echenard, \emph{Study of the radiation dose and
  neutron flux on the calorimeter},  Mu2e-doc-2853, 2013, unpublished.
  \bibitem{Hamamatsu} \verb"http://www.hamamatsu.com/us/en/index.html" 
\bibitem{FBK} \verb"https://srs.fbk.eu/optimization-sipm-technology "
  \bibitem{CALLIOPE}
  S.\ Baccaro, A.\ Cecilia and A.\ Pasquali, \emph{$\gamma$ irradiation facility 
  at ENEA-Casaccia Centre, Roma}, ENEA Report RT/2005/28/FIS (2005).
\bibitem{FNG}
  M.\ Martone. M.\ Angelone and M.\ Pillon, \emph{The 14 MeV Frascati neutron
  Generator}, Journal of Nuclear Materials 212-215 (1994) 1662-1664.
\end{thebibliography}
\end{document}